\documentclass[12pt,a4paper]{article}
\pdfoutput=1
\usepackage{color}
\usepackage{amssymb,amsmath,bm,bbm}
\usepackage{epsf}
\usepackage{epsfig}
\usepackage{afterpage}
\usepackage{longtable}
\usepackage[dvipsnames]{xcolor}
\usepackage[linktoc=page,bookmarks=false,colorlinks=false,
linkbordercolor=RoyalBlue,citebordercolor=ForestGreen,urlbordercolor=CornflowerBlue]{hyperref}
\usepackage{latexsym,mathrsfs,dsfont}
\usepackage[normalem]{ulem} 
\usepackage[compress]{cite}
\usepackage{graphicx}
\usepackage{url}
\usepackage{paralist}
\usepackage{bbold}

\newcommand{\OmHatEff}{\hat\Omega_\text{eff}}

\newcommand{\ord}{\mathcal{O}}

\newcommand{\RE}{{\rm Re}}

\newcommand{\gev}{\, {\rm GeV}}
\newcommand{\mev}{\, {\rm MeV}}

\newcommand{\bsi}{B_6^{(1/2)}}
\newcommand{\bei}{B_8^{(3/2)}}

\def\epe{\varepsilon'/\varepsilon}
\newcommand{\beq}{\begin{equation}}
\newcommand{\eeq}{\end{equation}}
\newcommand{\be}{\begin{equation}}
\newcommand{\ee}{\end{equation}}
\newcommand{\bi}{\begin{itemize}}
\newcommand{\ei}{\end{itemize}}
\newcommand{\ba}{\begin{array}}
\newcommand{\ea}{\end{array}}
\newcommand{\beqa}{\begin{eqnarray}}
\newcommand{\eeqa}{\end{eqnarray}}
\newcommand{\bea}{\begin{eqnarray}}
\newcommand{\eea}{\end{eqnarray}}
\newcommand{\beqn}{\begin{eqnarray}}
\newcommand{\eeqn}{\end{eqnarray}}

\newcommand{\re}{{\rm Re}}
\newcommand{\im}{{\rm Im}}

\definecolor{red}{cmyk}{0,1,1,0.4}

\usepackage{fancyhdr}
\advance \headheight by 15.0truept       

\begin{document}


\vspace{-14mm}
\begin{flushright}
        {AJB-20-2}\\
CP3-20-8
\end{flushright}

\vspace{8mm}

\begin{center}
{\Large\bf
\boldmath{Isospin-Breaking in $\epe$: Impact of $\eta_0$ \\at the Dawn of the 2020s }}
\\[12mm]
{\bf \large  Andrzej~J.~Buras ${}^a$ and Jean-Marc G\'erard${}^b$ \\[0.8cm]}
{\small
${}^a$TUM Institute for Advanced Study, Lichtenbergstr.~2a, D-85748 Garching, Germany\\
Physik Department, TU M\"unchen, James-Franck-Stra{\ss}e, D-85748 Garching, Germany\\[2mm]
${}^b$ Centre for Cosmology,
Particle Physics and Phenomenology (CP3), Universit{\'e} catholique de Louvain,
Chemin du Cyclotron 2,
B-1348 Louvain-la-Neuve, Belgium}
\end{center}

\vspace{8mm}

\begin{abstract}
  \noindent
  For direct CP-violation in $K\to\pi\pi$ decays, the usual isospin-breaking effects at the percent level are amplified by the dynamics behind  the $\Delta I=1/2$ rule and conventionally encoded in $\Omega_{\rm IB}$ parameters.
   The updated prediction  $\Omega_{\rm IB}^{(8)}=(15.9\pm 8.2)\times 10^{-2}$ 
of the Chiral Perturbation Theory  for 
the strong isospin-breaking due to $\pi_3-\eta_8$ mixing  confirms such a tendency but is quite sensitive to the theoretical input value of the low-energy constant corresponding to 
the flavour-singlet $\eta_0$ exchange contribution in this truncated octet scheme. We rather exploit the phenomenological $\eta_8-\eta_0$ mixing as a probe for the non-negligible flavour-singlet component of the physical $\eta$ pole
to find $\Omega_{\rm IB}^{(9)}=(35\pm7)\times 10^{-2}$  in a complete nonet scheme.
A large central value in the nonet scheme is thus substituted for a large uncertainty in the octet one.  Including the experimental $\pi^+-\pi^0$ mass difference as the dominant electromagnetic isospin-breaking, we obtain for the effective parameter entering the ratio $\epe$ an improved result  $\hat\Omega_{\rm eff}^{(9)}=(29\pm7)\times 10^{-2}$ to be   compared with
$\hat\Omega_{\rm eff}^{(8)}=(17\pm9)\times 10^{-2}$ used   in  recent analyses of $\epe$.
Accordingly, we get a reduction from
$(\epe)_\text{SM}^{(8)}=(17.4\pm 6.1)\times 10^{-4}$
to $(\epe)_\text{SM}^{(9)}=(13.9\pm 5.2)\times 10^{-4}$ and thereby an effective suppression of $(\epe)_\text{SM}$  by isospin-breaking corrections as large as $40\%$ relative to the recent RBC-UKQCD value.
  
\end{abstract}

\setcounter{page}{0}
\thispagestyle{empty}
\newpage

\tableofcontents

\section{Introduction}
During the last five years there has been  a renewed interest in the calculation
of the ratio $\epe$ \cite{Buras:2015yba,Kitahara:2016nld,Cerda-Sevilla:2016yzo, Cerda-Sevilla:2018hjk} in the Standard
Model (SM), after first results on the hadronic matrix elements from RBC-UKQCD lattice collaboration (LQCD) \cite{Bai:2015nea,Blum:2015ywa} that
hinted a possible anomaly. While this possibility  was soon supported by the Dual QCD (DQCD) approach  \cite{Buras:2015xba, Buras:2016fys}\footnote{For a most recent detailed description of this approach, see \cite{Buras:2014maa}.}, a different view
has been presented by the authors of  \cite{Gisbert:2017vvj,Cirigliano:2019cpi,Cirigliano:2019ani} who using
Chiral Perturbation Theory (ChPT) methods concluded that the Standard Model (SM) value
of $\epe$ agrees well with the world average
from NA48 \cite{Batley:2002gn} and KTeV
\cite{AlaviHarati:2002ye,Abouzaid:2010ny} collaborations 
\be\label{EXP}
(\epe)_\text{exp}=(16.6\pm 2.3)\times 10^{-4} \,.
\ee

But the 2015-2016 results of the RBC-UKQCD collaboration and of the DQCD team suffered from 
large uncertainties in hadronic matrix elements of
the dominant QCD penguin  (QCDP) operator $Q_6$ and still unsatisfactory
treatment of final state interactions. On the other hand, ChPT approach
encounters 
difficulties in matching long distance and short distance contributions.
Consequently, it was not possible to decide 
whether the hinted anomaly is really present or not.

The most recent and very much improved RBC-UKQCD calculation resulted
in \cite{Abbott:2020hxn}
\be\label{LAT}
(\epe)_\text{SM}=(21.7\pm 8.4)\times 10^{-4} \,,\qquad (\text{LQCD}-2020)
\ee
in agreement now with the experimental value in  (\ref{EXP}). Unfortunately, an  error of $39\%$ does not allow yet for clear
cut conclusions whether some amount of new physics contributions is present
in $\epe$ or not. The same applies to  
the earlier
updated ChPT analysis \cite{Cirigliano:2019cpi}
which resulted in
\begin{align}
  \label{Pich}
  (\epe)_\text{SM}  & = (14 \pm 5) \times 10^{-4} \,,\qquad (\text{ChPT}-2019),
\end{align}
with an error of $36\%$ that is very close to the LQCD one. { But it should 
  be remarked that with the present best values of the CKM parameters
  the central value in (\ref{Pich}) would be raised to $15.0\times 10^{-4}$.}

 If the experimental value (\ref{EXP}) is eventually confirmed by further numerical LQCD results, the analytical bound
  \be\label{DQCD}
  (\epe)_\text{SM}< (6.0\pm 2.4)\times 10^{-4}, \qquad (\text{DQCD}-2016)
  \ee
  derived in \cite{Buras:2015xba, Buras:2016fys} would then imply sizeable subleading contributions to the $Q_6$ hadronic matrix elements below $1~\gev$  in the DQCD approach.
However, the new RBC-UKCD result has been obtained by raising the matching scale from $1.5\gev$ to $4\gev$ in order to reduce the systematic error. As a consequence, this makes it difficult if not impossible for us to confront the numerical result (\ref{LAT}) with the analytical ones displayed in (\ref{Pich}) and
(\ref{DQCD}). Indeed, the LQCD approach is purely based on the quark and gluon dynamics with no reference to the meson degrees of freedom at work in both the ChPT and DQCD ones. Comments on both hadronic matrix elements and final state
interactions will be given later on. Meanwhile  we note that the excited-state contamination seen at $1.5\gev$ in LQCD data might be residual effects of the scalar resonance at
  $\ord(p^2,0)$ and meson evolution at $\ord(p^0,1/N)$, those being essential for the estimate of the $Q_6$ hadronic matrix elements around $1~\gev$ in DQCD.

{ Now isospin-breaking effects
 as well as NNLO QCD corrections to Wilson coefficients
of penguin operators, all absent in the RBC-UKQCD
result quoted above, lower the SM predition for $\epe$ \cite{Aebischer:2019mtr}.}
Extracting the hadronic matrix elements from \cite{Abbott:2020hxn},
 using the estimate
of  the isospin-breaking  effects from \cite{Cirigliano:2019cpi} and including the missing NNLO QCD corrections to electroweak penguin contributions, 
the authors in {\cite{Aebischer:2020jto}} find
  \be\label{6.10ba}
  (\epe)_{\rm SM}=(17.4\pm 6.1)\times 10^{-4}\,,
  \ee
  in a very good agreement with experiment and the expectations in (\ref{Pich}).
  { Still uncertainties are not in full theoretical control: discrepancies
    of $\ord(10^{-3})$ between the experimental value and the SM prediction
    could be either positive, as expected from DQCD, or negative as now allowed by the new RBC-UKQCD result.} 

  As analyzed in detail in \cite{Aebischer:2019mtr,Aebischer:2020jto}, the main
  left-over uncertainties in the evaluation of $\epe$ reside in the
  hadronic matrix element of the QCDP operator $Q_6$ and the isospin-breaking (IB) effects.   It is expected that future lattice
  calculations will reduce the errors on the corresponding $\bsi$ and $\hat\Omega_{\rm eff}$ parameters,
  but   this could still take several years. The goal of our paper is
  to demonstrate that a better estimate for the strong and electromagnetic components of $\hat\Omega_{\rm eff}$ can
  already be done at the dawn of the 2020s.

  Our paper is organized as follows.
  In Section~\ref{sec:1}, after recalling how  the parameter $\hat\Omega_{\rm eff}$ enters
  the basic formula for $\epe$, we summarize the results for
  its strong IB component $\Omega_{\rm IB}^{(8)}$ recently obtained using ChPT in
  the octet scheme for light pseudoscalars and   $\Omega_{\rm IB}^{(9)}$ obtained by us in 1987 rather
  employing a simple $\eta-\eta^\prime$ phenomenological model (PhM)   \cite{Buras:1987wc} in a nonet scheme. {In the latter scheme the impact  of the flavour-singlet   meson $\eta_0$ on $\epe$ is explicitly included while in the octet
  scheme, necessarily used in ChPT, it is  buried in a poorly known low-energy constant.}

  In Section~\ref{sec:2} we update the calculation of  $\Omega_{\rm IB}^{(9)}$
  in PhM finding that, with better known input parameters, its value increases
  by $25\%$ relative to our 1987 result and is thereby significantly larger
  than the central value of $\Omega_{\rm IB}^{(8)}$ from ChPT.
  
  In Section~\ref{sec:3}  we  make use of the minimal chiral theory at $\ord(p^0, 1/N) + \ord(p^2, 0)$, N being the number of colours, to validate the update of  $\Omega_{\rm IB}^{(9)}$ and highlight the impact of the flavour-singlet
 $\eta_0$ on the 
  evaluation of $\epe$ thereby questioning the octet
  scheme in this context.

  In  Section~\ref{sec:4} we derive the formula for $\Omega_{\rm IB}^{(9)}$
  as a function of the $\eta_8-\eta_0$ mixing angle which allows us to compare 
  transparently strong isospin-breaking in the nonet scheme with the one
  in the octet scheme. This comparison is complemented in Section~\ref{sec:5}
  with the help of a Single Resonance Approximation to  ChPT \cite{Ecker:1988te}.

  In  Section~\ref{sec:6} we summarize our results for $\Omega_{\rm IB}^{(9)}$,
  add to it the dominant electromagnetic IB component to complete
   $\hat\Omega^{(9)}_{\rm eff}$, comment on the { controversial} scale dependence of $\bsi$ and eventually present our estimate for $\epe$ that
  updates the one based on $\hat\Omega^{(8)}_{\rm eff}$ in {\cite{Aebischer:2020jto}}.
  Brief conclusions about our main findings  are given in  Section~\ref{sec:7}.

\boldmath
\section{Isospin-breaking impact on $\epe$}\label{sec:1}
\unboldmath
In order to understand the way $\hat\Omega_{\rm eff}$ enters the evaluation of
$\epe$,
we simply recall that the basic formula 
\be
  \label{eprime0}
\epe=\frac{1}{\sqrt{2}\,|\varepsilon_K|}{\im}\left(\frac{A_2}{A_0}\right)= -\frac{\omega}{\sqrt{2}\,|\varepsilon_K|}\frac{\im A_0}{\re A_0}\left(1-\frac{1}{\omega}\frac{\im A_2}{\im A_0}\right),
    \ee
    implies an $\omega^{-1}\approx 22$ amplification by the $\Delta I=1/2$ rule of any CP-violating contribution to the isospin $I=2$ amplitude $A_2$ in $K^0\to\pi\pi$ {decays \cite{Donoghue:1986nm}.} This $\Delta I=1/2$ rule is dominated by the QCD dynamics acting on current-current operators, as already found many years ago within DQCD \cite{Bardeen:1986vz,Buras:2014maa} 
    and also seen recently in LQCD data \cite{Blum:2015ywa,Abbott:2020hxn,Donini:2020qfu}.
    
     Now within the CKM paradigm the dominant contributions to ${\im A_2}$ originate from the electroweak
    penguin (EWP) contributions  and from  isospin-breaking (IB) effects.
    Neglecting subleading effects which
    will be discussed in Section~\ref{sec:6},
    we thus decompose  ${\im A_2}$ without any ambiguity 
    \be\label{DECOMP}
       {\im A_2} \approx  ({\im A_2})^{\rm EWP}+ ({\im A_2})^{\rm IB}\,
       \ee
to rewrite (\ref{eprime0}) as follows
\begin{align}
  \label{eprime}
  \epe &
  = -\,\frac{\omega}{\sqrt{2}\,|\varepsilon_K|}
  \left[\, \frac{\im A_0}{\re A_0}\, (1 - \OmHatEff)
    -    \, \frac{(\im A_2)^{\rm EWP}}{\re A_2} \,\right],\quad
  \hat\Omega_{\rm eff}\approx \Omega_\text{IB}=\frac{1}{\omega}\frac{(\im A_2)^{\rm IB}}{\im A_0}\,.
\end{align}

The amplitudes ${\re A_0}$ and ${\re A_2}$ in this formula are extracted from the data in the isospin limit. Our $\OmHatEff$ differs from $\Omega_\text{eff}$ in \cite{Cirigliano:2003nn, Cirigliano:2003gt,Cirigliano:2019cpi}
as it does not include EWP contributions to
$\im A_0$ summarized in these papers by $\Delta_0$. 
We find it more natural indeed to
calculate $\im A_0$ including both QCDP and EWP contributions as this allows to
keep track of any NP contributions to $\im A_0$. In fact the RBC-UKQCD collaboration
includes EWP contributions to $\im A_0$ as well.

The impact of  strong isospin-breaking induced by the $(m_u-m_d)$ quark mass difference  on the direct CP-violating parameter $\epe$ has been recently revisited in \cite{Cirigliano:2019cpi} in the context of ChPT. Within the {\em octet scheme} of this approach, the estimate
 at the next-to-leading $\ord(p^4)$  level 
\be\label{ChPT}
  \Omega_{\rm IB}^{(8)}=(15.9\pm 8.2)\times 10^{-2}\quad {\rm at} \quad \ord(p^4)\qquad
    (\rm{ChPT}-2019)\,,
  \ee
 amounts to  a small increase compared to the leading $\ord(p^2)$ one:
\be\label{ChPTa}
\Omega_{\rm IB}^{(8)}=13.8 \times 10^{-2}\quad {\rm at} \quad \ord(p^2)\,.
  \ee

  In the absence of electromagnetic isospin-breaking induced for its part by the
  $(q_u-q_d)$ quark charge difference, this trend confirms an earlier estimate made within the same framework \cite{Cirigliano:2003gt}, i.e., $(15.9\pm 4.5)\times 10^{-2}$. Yet, with its larger uncertainties to be explained later, the result (\ref{ChPT}) is
  also not inconsistent (at the $1.4\,\sigma$ level) with the value obtained already in 1987 in a phenomenological model (PhM) including the $\eta^\prime$  in a {\em nonet scheme} \cite{Buras:1987wc}
\be\label{PhM}
  \Omega_{\rm IB}^{(9)}\approx 27\times 10^{-2} ,\qquad
    (\rm{PhM}-1987)\,.
  \ee

  We will next provide an update of the latter approach. Leading to a $25\%$ increase for the numerical value quoted in (\ref{PhM}) above, it implies that
    the difference between the octet and nonet estimates is significantly
    larger than the comparison of the 2019 result in (\ref{ChPT}) and 1987 result in (\ref{PhM}) would indicate.

\boldmath
\section{Update of a simple $\eta-\eta^\prime$ pole model}\label{sec:2}
  \unboldmath
  Let us assume
  a single-pole dominance of the two lowest-lying pseudoscalar isosinglets
  \be\label{2.1}
  \eta\equiv \eta_8\cos\theta-\eta_0\sin\theta, \qquad \eta^\prime\equiv \eta_8\sin\theta+\eta_0\cos\theta,
  \ee
  that mix with the $\pi_3$ component of the pion isotriplet in $K^0\to\pi^0\pi^0$ decay. Under this assumption, we easily recover the first order strong IB correction to $\epe$  \cite{Buras:1987wc}:
  \be\label{2.2}
  \Omega_{\rm IB}^{(9)}=\left(\frac{4}{9\sqrt{2}\omega R}\right)(m_K^2-m_\pi^2)
  \left[\frac{(\cos\theta-\sqrt{2}\sin\theta)^2}{(m_\eta^2-m_\pi^2)}+
    \frac{(\sin\theta+\sqrt{2}\cos\theta)^2}{(m_{\eta^\prime}^2-m_\pi^2)}\right]\,.
  \ee
  In particular, the $\sqrt{2}$ factors in the numerators arise from the flavour-singlet $\eta_0$ component of
  $\eta$ and $\eta^\prime$, as seen in (\ref{2.1}). They result from the relative
off-shell matrix elements at $\ord(p^2)$
   \be\label{JMG1}
  \frac{\langle\pi\eta_0|Q_6|K\rangle}{\langle\pi\eta_8|Q_6|K\rangle}=\sqrt{2},
  \ee and 
  subsequent IB mixings
  \be\label{JMG2}
  \frac{\langle\pi^0|\eta_0\rangle}{\langle\pi^0|\eta_8\rangle}=\sqrt{2}.
  \ee
  
In  (\ref{2.2}), the famous $\Delta I = 1/2$ rule enhancement factor in the
  $K\to\pi\pi$ decay amplitudes  \cite{GellMann:1955jx,GellMann:1957wh}
  \be\label{Delta1/2}
  \frac{1}{\omega}=\frac{{\RE}A_0}{{\RE}A_2}=22.36\pm 0.05\,,
  \ee
  as determined in \cite{Cirigliano:2019cpi},
  (almost) balances the smallness of the SU(2) breaking compared to the SU(3) one \cite{Aoki:2019cca}:
  \be\label{2.4}
  R\equiv \frac{[m_s-(m_u+m_d)/2]}{(m_d-m_u)}=38.1\pm 1.5\,.
  \ee

  For the physical pseudoscalar masses in (\ref{2.2}), we take the central values \cite{Tanabashi:2018oca}
  \be\label{2.5}
  m_\pi=0.135\gev,\qquad m_\eta=0.5479\gev,\qquad m_{\eta^\prime}=0.9578\gev\,,
  \ee
  as well as the convention 
  \be\label{2.6}
  m_K^2\equiv \frac{(m_{K^0}^2+m_{K^+}^2)}{2}=(0.4957\gev)^2\,.
  \ee

  In this phenomenological framework, the $\eta_8-\eta_0$ mixing angle $\theta$ is a free parameter independent of the $\eta-\eta^\prime$  mass spectrum. So, we assume (for the time being) the following phenomenological value for this angle
  \be\label{2.7}
  \theta_{\rm ph}=-19.47^\circ\,, \quad \tan\theta_{\rm ph}\equiv-\frac{1}{2\sqrt{2}}
  \ee
  which turns out to be quite compatible with the complete set of
$J/\psi\to\gamma(\pi,\eta,\eta^\prime)$ 
  branching ratios \cite{Gerard:2013gya}. Corresponding to $\cos\theta_{\rm ph} = 2\sqrt{2}/3$ and
  $\sin\theta_{\rm ph} = -1/3$, this mixing angle implies the mnemonic wave-functions
  \be\label{2.8}
\eta\equiv \frac{1}{\sqrt{3}}(\bar u\gamma_5u+\bar d\gamma_5 d-\bar s\gamma_5 s),\qquad 
\eta^\prime\equiv \frac{1}{\sqrt{6}}(\bar u\gamma_5u+\bar d\gamma_5 d+2\bar s\gamma_5 s)
\ee
in the framework of a naive quark model.

As seen from (\ref{2.8}), the peculiar rotation angle (\ref{2.7}) simply amounts to identify $\eta$ to $\eta_0$ and $\eta^\prime$ to $\eta_8$ 
up to an $s\bar s$ flip of sign in their hadronic matrix elements and so easily explains
the huge suppression of $B^0\to K^0\eta$ 
relative to $B^0\to K^0\eta^\prime$ on-shell transitions
through the dominant penguin operator
$(\bar b_L d_R)(\bar d_R s_L)+(\bar b_L s_R)(\bar s_R s_L)$ \cite{Gerard:2006ch}.
Similarly, the resulting interchange of renormalization factors when going from the octet-singlet flavour basis $(\eta_8,\eta_0)$ to the nonet one in (\ref{2.8})  implies a switch in (\ref{2.2}) of the $(\sqrt{2})^2$ enhancement factor from $\eta_0$ (for $\theta=0$) to $\eta$ (for $\theta=\theta_\text{ph}$) for
$K^0\to \pi^0\pi^0$ transitions via {\em off-shell}
$Q_6$ matrix elements. 
From (\ref{2.2})-(\ref{2.7}), we indeed get 
\be\label{2.9}
\Omega_{\rm IB}^{(9)}=0.138\left(\frac{4(m_K^2-m_\pi^2)}{3(m_\eta^2-m_\pi^2)}\right)
\left(2+\frac{(m_\eta^2-m_\pi^2)}{(m_{\eta^\prime}^2-m_\pi^2)}\right)=
34\times 10^{-2}\qquad 
 (\rm{PhM}-2020)\,
  \ee
  which is now more than $2\sigma$  above the central value (\ref{ChPT}) of ChPT. We emphasize that the lower numerical result obtained in \cite{Buras:1987wc} and recalled in   (\ref{PhM}) mostly originates from
  the input values available at that time for the parameters in (\ref{2.4}) and  (\ref{2.7}), namely $R = 45.5$ and $\theta = -22^\circ$, respectively.

  In this phenomenological approach, uncertainties on (\ref{2.9}) are basically due to our assumption expressed at the begining of this section and seem difficult to assess.   The pseudoscalar pole dominance
   below one $\gev$ will be discussed in more details in Section~\ref{sec:5}.

So, let us now turn to a framework where the $\eta_8-\eta_0$ mixing angle $\theta$ is not free anymore but fixed by the $\eta-\eta^\prime$ mass spectrum to assign
theoretical uncertainties on (\ref{2.9}).

\boldmath
\section{Correlation between  $\eta^{(\prime)}$ masses and mixing}\label{sec:3}
\unboldmath
Any theoretical framework   allowing us to  connect the octet scheme to the nonet one in a simple way is more than welcome in order to directly confront the ChPT result (\ref{ChPT}) with the updated PhM value (\ref{2.9}).
For that purpose,  we consider the complete effective Lagrangian at $\ord(p^2,0)+ \ord(p^0,1/N)$:
\be\label{3.1}
L=\frac{F^2}{8}\left[\langle\partial_\mu U\partial^\mu U^+\rangle +r \langle m U⁺+Um^+\rangle+\frac{m_0^2}{12}\langle \ln U-\ln U^+\rangle^2\right]
\ee
with
\be\label{3a}
U=\exp(i\sqrt{2}\pi/F), \qquad \pi=\lambda^a\pi_a~ (a=0,...8)
\ee
the unitary chiral matrix for the nonet of light pseudoscalars. In our notations, the $F$ and $r$ scale parameters in (\ref{3.1}) are related to the pion weak decay constant and mass via the relations
\be\label{Fr}
F\approx F_\pi=130\mev, \qquad r=\frac{2 m_\pi^2}{m_u+m_d},
\ee
while the $m_0$ one provides the $\eta_0$ with a large anomalous mass to solve the so-called
$\text{U(1)}_\text{A}$ problem, namely to insure $m^2_{\eta^\prime}\gg m^2_{\eta}$.

On the basis of (\ref{3.1}), the $\eta_8-\eta_0$ mixing angle is not free anymore. The theoretical isosinglet square masses, consistently considered in the isospin limit for the first order correction (\ref{2.2}), are indeed given by the relations
\be\label{3.3a}
m_\eta^2=\frac{1}{3}[(4 m_K^2-m_\pi^2)+2\sqrt{2}(m_K^2-m_\pi^2)\tan\theta],
\ee

\be\label{3.3b}
m_{\eta^\prime}^2=\frac{1}{3}[(4 m_K^2-m_\pi^2)-2\sqrt{2}(m_K^2-m_\pi^2)\cot\theta],
\ee
with $\theta\in[-\pi/4,+\pi/4]$. The ``ideal'' $\theta=\tan^{-1}(1/\sqrt{2})\approx +35.26^\circ$ angle corresponds to $m_0=0$ with $m_{\eta^\prime}^2=m_\pi^2$, as it
should. Yet, two distinct values of the mixing angle 
\be\label{3.4}
\theta_\eta= -5.68^\circ, \quad {\rm and} \quad \theta_{\eta^\prime}= -19.80^\circ
\ee
are extracted from (\ref{3.3a}) and (\ref{3.3b})
respectively, if the physical masses given in (\ref{2.5}) and (\ref{2.6}) are again used as inputs. At the source of this seeming clash, one finds the ratio
\be\label{3.5}
\Delta\equiv\frac{(m_\eta^2-m_\pi^2)}{(m_{\eta^\prime}^2-m_\pi^2)}
\ee
that cannot be reproduced whatever the value of a single $\theta$ angle
\cite{Georgi:1993jn,Gerard:2004gx}.
From (\ref{3.3a}) and  (\ref{3.3b})   we indeed get:  

\be\label{Delta}
\Delta=\Delta(\theta)= \tan\theta \times \left(\frac{\tan\theta+\sqrt{2}}{\sqrt{2}\tan\theta-1}\right)\equiv\tan\theta\times\tan(2\delta-\theta), \qquad \tan 2\delta=-\sqrt{2}
\ee
with a theoretical upper bound for $\theta=\delta\approx-27.4^\circ$,
namely $\Delta(\theta)<\tan^2(\delta)=2-\sqrt{3} = 0.268$
, at variance by about $20\%$  with  the experimental value derived from the physical masses in (\ref{2.5}), i.e., $\Delta=0.314$. So, the minimal effective Lagrangian (\ref{3.1}) is over-constraining compared to the phenomenological approach used in Section~\ref{sec:2}. Higher-order terms are in principle required to reconcile the two mixing angles displayed in (\ref{3.4}) and, simultaneously, to reproduce the observed mass ratio $\Delta$  defined in (\ref{3.5}). But such a tedious approach involving further hadronic mass scale parameters at subleading 
$\ord(p^4,0)+ \ord(p^2,1/N)+\ord(p^0,1/N^2)$ in our scheme
can be avoided if, inspired by (\ref{3.4}), one first scrutinizes two limits for the over-constrained mass relations (\ref{3.3a}) and (\ref{3.3b}). They will eventually help us connecting the octet scheme of ChPT with the nonet one of PhM in a simple way, our main purpose after all.

On the one hand, in the limit of vanishing $\eta_8-\eta_0$ mixing (as suggested by $\theta_\eta$), we consistently recover the pure octet scheme with its surprisingly successful Gell-Mann-Okubo (GMO) mass relation for the $\eta(548)$ and a full decoupling of the $\eta^\prime$:
\be\label{3.7}
m_\eta^2=m_8^2\equiv\frac{1}{3}(4 m_K^2-m_\pi^2)=(0.567 \gev)^2, \qquad(\theta=0),\qquad
m_{\eta^\prime}^2=m_0^2=\infty\,.
\ee

On the other hand, in the limit of phenomenological $\eta_8-\eta_0$ mixing
(\ref{2.7}) (as suggested by $\theta_{\eta^\prime}$),
we obtain a (by far) more realistic spectrum with a quite successful mass relation for the $\eta^\prime(958)$ this time and a still reasonable value for the $\eta$ mass:
\be\label{3.8}
m_\eta^2=m_K^2=(0.496\gev)^2, \quad (\theta_{\rm ph}=-19.5^\circ), \quad
m_{\eta^\prime}^2=4m_K^2-3 m_\pi^2=(0.963\gev)^2.
\ee

Contrary to the theoretical $\eta^\prime$ mass in (\ref{3.3b}), the $\eta$ one in (\ref{3.3a}) displays a very weak dependence on the mixing angle (see Fig.\,1 in \cite{Degrande:2009ps}). As a matter of fact, $m_\eta$ fluctuates around its physical value from $4\%$ up in (\ref{3.7}) to $-10\%$ down in (\ref{3.8}) for $\theta$
in the range $[0,\theta_{\rm ph}]$. 
From this perspective, we conclude that the success of the GMO mass relation at the $\ord(p^2, 0)$ level looks somewhat accidental. Just for comparison, the observed splitting of the K and $\pi$ weak decay constants is, as for the ratio $\Delta$ in (\ref{3.5}), of the order of $20\%$ 
above the theoretical degeneracy predicted by the effective Lagrangian
(\ref{3.1}):  
\be\label{3.9}
\frac{F_K}{F_\pi}=1.19\,.
\ee
Deviations at the level of $20\%$ are in fact expected from generic SU(3)-breaking corrections of order $(m^2_K -m^2_\pi)/\Lambda^2$, $\Lambda$ being a typical hadronic mass scale around one GeV. This is indeed the case through the next-to-leading $\ord(p^4, 0)$ term beyond our effective Lagrangian (\ref{3.1}):
\be\label{3.10}
\delta L=-\left(\frac{F^2}{8\Lambda^2}\right)\langle r m\partial^2U^++h.c.\rangle
\ee
or, equivalently (in the octet limit),
\be\label{3.11}
\delta L=\left(\frac{F^2}{8\Lambda^2}\right)
\langle r m U^+\partial_\mu U\partial^\mu U^++h.c.\rangle.
\ee
The usual $\partial_\mu \to D_\mu=\partial_\mu-igW_\mu$ substitution in the presence of weak gauge interactions leads to
\be\label{3.12}
\frac{F_K}{F_\pi}=1+\frac{(m_K^2-m_\pi^2)}{\Lambda^2}
\ee
in the large N limit 
(i.e., in the absence of the 1/N-suppressed chiral loop contributions).
From (\ref{3.9}) and (\ref{3.12}), we easily infer a hadronic mass scale in the ballpark of observed scalar resonances:
\be
1\gev < \Lambda\approx 1.1\gev < 1.5\gev
\ee 
as suggested by a linear $\sigma$-model in the large N limit
\cite{Gerard:1990dx}.
 In fact, the same scalar resonance effect governs our estimate of the $Q_6$ hadronic matrix elements at $\ord(p^2,0)$  \cite{Bardeen:1986vp}. 

Let us now turn the argument the other way around. The mixing angle extracted from the mass relation (\ref{3.3a}) for $\eta$ , namely
\be\label{3.14}
\tan\theta_\eta=\left(\frac{3}{2\sqrt{2}}\right)\frac{(m^2_\eta-m_8^2)}{(m_K^2-m_\pi^2)},
    \ee
  is clearly quite sensitive to any deviation from the GMO mass relation. As a crucial consequence, it strongly depends on the higher-order terms beyond the effective Lagrangian (\ref{3.1}). For a simple illustration we consider again the next-to-leading $\ord(p^4, 0)$ term (\ref{3.11}) that would, alone, lead to the following shift for the GMO mass relation: 
  \be
  m_8^2\rightarrow \frac{1}{3} (4 m_K^2-m_\pi^2)-
  \left(\frac{8}{9}\right)\frac{(m_K^2-m_\pi^2)^2}{\Lambda^2}.
  \ee
  Yet, inserting this $6\%$ decrease of $m_8$  in
  (\ref{3.14}), we would then obtain a totally unrealistic $\eta_8-\eta_0$
  mixing angle shifted from $-6^\circ$ to $+4^\circ$ for $m_\eta\approx m_\eta^{\rm phys}$.

\boldmath
\section{Strong isospin-breaking: nonet vs. octet}\label{sec:4}
\unboldmath
As already stated, the over-constraint on the $\eta-\eta^\prime$  mass spectrum
inferred from (\ref{3.1}) is in fact welcome since the pole correction factor given in (\ref{2.9}) can now be theoretically expressed and dissected in terms of the mixing angle only, thanks to the relations (\ref{3.3a}), (\ref{3.3b}) and (\ref{Delta}):
\be\label{4.1}
\boxed{\Omega_{\rm IB}^{(9)}(\theta)=0.138\, \left(1+\frac{\tan\theta}{\sqrt{2}}\right)^{-1}
\left[(\cos\theta-\sqrt{2}\sin\theta)^2+(\sin\theta+\sqrt{2}\cos\theta)^2\Delta(\theta)\right].}
\ee

\begin{itemize}
\item
  If the $\theta$ angle is equal to $0^\circ$, we consistently recover the leading  result (\ref{ChPTa}), namely 
  \be\label{ChPTb}
  \Omega_{\rm IB}^{(9)}(0)=13.8\times 10^{-2}
  \ee
  with a full decoupling of $\eta^\prime$ (i.e., $\Delta = 0$).
\item
  If the $\theta$ angle is equal to $- 5.68^\circ$, as extracted from a very strong dependence on the $\eta$  mass in (\ref{3.3a}) (see Fig.\,1 of \cite{Degrande:2009ps}),
  we obtain 
    \be
  \Omega_{\rm IB}^{(9)}(\theta_\eta)=22.0\times 10^{-2}
  \ee
  with a totally unrealistic $\eta^\prime$ mass (i.e., 1.574 GeV).
\item
  If the $\theta$ angle is equal to $-19.8^\circ$, as extracted from a rather weak dependence on the $\eta^\prime$ mass in (\ref{3.3b}) (see Fig.\,1 of
  \cite{Degrande:2009ps}), we can safely consider its phenomenological value (\ref{2.7}) to display a quite simple anatomy of the strong IB parameter 
   \be
  \Omega_{\rm IB}^{(9)}(\theta_{\eta^\prime})\approx 0.138\left(\frac{4}{3}\right)[2+\Delta]  = 41.4\times 10^{-2}
  \ee
  namely a sizeable increase of (\ref{ChPTb}) due to a correction to the GMO mass relation (factor 4/3), a huge enhancement of the $\eta$ pole contribution by its flavour-singlet $\eta_0$ component (factor 2) and a very modest contribution from the heavier $\eta^\prime$ pole ($\Delta\approx 1/4$).
\end{itemize}
\section{Strong isospin-breaking: octet vs. nonet}\label{sec:5}
On the basis of a Single Resonance Approximation (SRA) to ChPT \cite{Ecker:1988te}, the authors of \cite{Ecker:1999kr} put forward a strong destructive interference from the $\ord(p^4)$ Lagrangian:
\be\label{5.1}
\delta L=L_7 \langle r m U^+-h.c\rangle^2+L_8 \langle rm U^+r m U^++h.c\rangle
+L_5 \langle rm U^+\partial_\mu U\partial^\mu U^++h.c\rangle
\ee
to explain the rather modest increase observed  when going from (\ref{ChPTa}) to
(\ref{ChPT})
in the octet scheme.
    In ChPT, the $L_i$ coefficients of (\ref{5.1}) are low-energy constants (LEC’s) that absorb the one-loop divergences at, say, the $\rho$ meson mass scale. As such, they encode the non-perturbative QCD effects. Taking care of the usual ChPT notations $B_0 = r/2$ and $\chi= rm$
    compared to ours and in accordance with our normalizations (\ref{Fr})
replacing
$F$ by $F/\sqrt{2}$  in ChPT formulae, we expect from (\ref{3.11}) all three LEC’s to be around $10^{-3}$. However, either improved data fits or further theoretical hypotheses have to be considered to fix them more precisely.
    In this SRA above the $\eta$ meson mass, only the effective $\ord(p^4, 1/N)$ $L_7$ term corresponding to the heavy pseudoscalar ($\eta_0$) tree-level exchange is renormalization scale independent, with the numerical value
    \be\label{37}
    L_7=-\frac{F_\pi^2}{96 m_0^2}\approx -0.3\times 10^{-3}\qquad 
    (\rm{ChPT}-1989)
    \ee
   if $m_0=\ord(0.8\gev$). Being fully contained in this $L_7$ LEC, the $\eta^\prime$ dominance effect taken alone in the {\em octet scheme} would then lead to a significant correction to the GMO mass relation
\be\label{GMO5}
m_8^2\rightarrow \left(\frac{1}{3}\right)(4 m_K^2 - m^2_\pi) + \left(\frac{256}{3F^2_\pi}\right) (m_K^2 - m_\pi^2)^2 L_7\approx m^2_K
\ee
that is quite consistent with our phenomenological limit (\ref{3.8}) in {\em the nonet scheme} (\ref{3.1}), as it should be. In fact, $m_8^2\rightarrow m_K^2$ if $L_7=-0.29\times 10^{-3}$, i.e., $m_0=0.78\gev$.
As a consequence, the SRA of ChPT provides a second theoretical  framework  allowing us to connect the octet scheme with the nonet one.

In the SRA of \cite{Ecker:1988te} the early input values for $L_8$ and $L_5$ LECs in
(\ref{5.1}) were
\be\label{5.5}
L_8^r(m_\rho)\approx 0.9\times 10^{-3},\qquad L_5^r(m_\rho) \approx 1.4\times 10^{-3}, \qquad     (\rm{ChPT}-1989)\,.
\ee
The $L_8$ term of (\ref{5.1}) involves (like the $L_5$ one) tree-level scalar exchange and its positive value in (\ref{5.5}) turns out to cancel the $L_7$ correction in the linear combination $(3L_7 + L_8)$ that precisely enters the
$\ord(p^4)$ correction to  the $\pi_3-\eta_8$ mixing \cite{Gasser:1984gg}. According to \cite{Ecker:1999kr}, this full destructive interference would explain the rather mild $\ord(p^4)$ correction to $\Omega_{\rm IB}^{(8)}$, as still displayed in (\ref{ChPT}). Interestingly such a strong destructive interference simultaneously occurs for another   linear combination of LECs
that enters the complete  $\ord(p^4)$ correction to the GMO mass relation
(\ref{GMO5}) this time, namely $(L_7 + L_8/2-L_5/12)$.

Is it then the final argument in favour of the relatively stable value displayed in (\ref{ChPT}) compared to (\ref{ChPTa})?
Well, not really since this heuristic interpretation in terms of accidental destructive interferences among tree-level single resonance contributions have to be taken with a grain of salt. Indeed, the nature and mass spectrum of the low-lying scalar states are still rather controversial nowadays with, in particular, the broad $f_0(500)$ resonance \cite{Tanabashi:2018oca}. Consequently, and contrary to $L_7$, the $L^r_8$ and $L^r_5$ LECs cannot be considered (yet) as evidence for scalar meson dominance \cite{Pich:2018ltt}. As a matter of fact, the early input values for $L^r_8$ and $L_5^r$ LECs given in (\ref{5.5})
were based on the identification of the scalar mass with the light $a_0 (980)$ one. However, their more precise values adopted in the numerical analysis of  \cite{Cirigliano:2019cpi},  namely 
\be\label{5.6}
L_8^r(m_\rho) = (0.53\pm 0.11) \times 10^{-3},\quad
L_5^r(m_\rho) = (1.20\pm0.10) \times 10^{-3},     (\rm{ChPT}-2019)
\ee
mostly rely on new LQCD results and rather favour the identification of the scalar mass with the heavier $a_0 (1450)$ one. This theoretical move of the relevant scalar mass scale from $\ord(1\gev)$ to $\ord(1.5\gev)$ is of course legitimate but has a sizeable suppression effect on the $L_8$ LEC. Doing so, it clearly invalidates the heuristic SRA argument put forward in \cite{Ecker:1999kr} since  only a partial destructive interference is actually at work in the linear combination $(3L_7 + L_8)$. As a consequence, the updated result (\ref{ChPT}) is now very sensitive to the input value of $L_7$ which (contrary to $L_{5,8}$) only relies on recent ChPT data fits
\be\label{42}
L_7=-(0.32\pm 0.10)\times 10^{-3},\qquad (\rm{ChPT}-2019).
\ee
This  eventually explains why the theoretical uncertainties quoted in the introduction for $\Omega_{\rm IB}^{(8)}$
are larger in  \cite{Cirigliano:2019cpi} than previously in \cite{Cirigliano:2003gt}:
with its $\eta_0$-dominance, the SRA backfires on the octet scheme.

Therefore, the main message of this section is that one should try as much as possible to avoid the somewhat accidental GMO mass relation for the pseudoscalar $\eta$. In other words, it is our opinion that any octet scheme leading to a $\Omega_{\rm IB}^{(8)}$ is not appropriate for the study of the strong isospin violation effect on 
$\epe$. As a matter of fact, physical processes clearly favour $\theta_{\eta^\prime}$ from (\ref{3.3b}) over $\theta_\eta$ from (\ref{3.3a}), as seen in (\ref{3.4}).

\boldmath
\section{ Summary on $\Omega_{\rm IB}$ and its impact on $\epe$}\label{sec:6}
\unboldmath
Within a nonet scheme, we have seen that $\Omega_{\rm IB}^{(9)}$ can be expressed in terms of  the $\eta^{(\prime)}$ masses and mixing through the equation
 (\ref{4.1}). If we privilege the physical masses in (\ref{2.5}) over the angle $\theta$, we then face the following alternative as given in (\ref{3.4}), that is either 
\be
m_\eta=m_\eta^{\rm phys}; \quad m_{\eta^\prime}=1.64\, m_{\eta^\prime}^{\rm phys}
\Rightarrow \Omega_{\rm IB}^{(9)}(\theta=-5.68^\circ)=22.0\times 10^{-2} \,,
\ee
or
  \be
m_{\eta^\prime}= m_{\eta^\prime}^{\rm phys}; \quad m_{\eta}=0.90\, m_{\eta}^{\rm phys} \Rightarrow \Omega_{\rm IB}^{(9)}(\theta=-19.80^\circ)=41.9 \times 10^{-2},
\ee
while the $\eta-\eta^\prime$ square mass ratio $\Delta$  defined in (\ref{3.5}) is optimized for $\Omega_{\rm IB}^{(9)}(\theta=-27.4^\circ)= 0.55$. However, our confrontation of the GMO mass relation  with the SRA of ChPT in the previous section prompts us to rather privilege more realistic values for the $\eta_8-\eta_0$ mixing angle, say
\be\label{6.2}
-20^\circ < \theta < - 10^\circ,
\ee
over specific $\eta$ and $\eta^\prime$ masses. Doing so in the nonet scheme
(\ref{3.1}), we obtain respectively
\be
\Omega_{\rm IB}^{(9)}(\theta=-10^\circ)=28\times 10^{-2} \Rightarrow
m_\eta=0.97\, m_\eta^{\rm phys}; \quad m_{\eta^\prime}=1.30\, m_{\eta^\prime}^{\rm phys}
\ee
\be
\Omega_{\rm IB}^{(9)}(\theta=-20^\circ)=42\times 10^{-2} \Rightarrow
m_\eta=0.90\, m_\eta^{\rm phys}; \quad m_{\eta^\prime}=1.00\, m_{\eta^\prime}^{\rm phys}
\ee
together with $\Omega_{\rm IB}^{(9)}(\theta=-18.4^\circ)=0.40$
for $(m_\eta^2+m_{\eta^\prime}^2)=(m_\eta^2+m_{\eta^\prime}^2)^{\rm phys}$.
The range  allowed for the  mixing angle in (\ref{6.2})
thus implies the theoretical estimate
  \be\label{6.4}
  \boxed{\Omega_{\rm IB}^{(9)}=(35\pm7)\times 10^{-2}}
  \ee
  that essentially amounts to assign a $20\%$ uncertainty on the updated PhM result (\ref{2.9}), as it was already the case for the square mass ratio
  $\Delta(\theta_{\rm ph})$ defined by (\ref{Delta}). A large central value in the nonet scheme is thus substituted for a large uncertainty in the octet one.

  As far as the strong isospin-breaking effect on the $A_2$  amplitude is concerned, only the upper side of (\ref{ChPT}) is clearly favoured by our nonet scheme, leading in principle to a sizeable decrease of the predicted value for $\epe$ within the octet one. However, electromagnetic  corrections should also be considered before drawing any firm conclusion since
   \be\label{49}
    \hat\Omega_{\rm eff}=\Omega_{\rm IB}^{(\rm strong)}+\Omega_{\rm IB}^{(\rm em)}.
    \ee
    The $\Omega_{\rm IB}^{(\rm strong)}$ component in (\ref{49}) corresponds to the strong isospin-breaking effects ($m_u\not=m_d$) on the dominant $Q_{4,6}$ {\em off-shell} matrix elements  via 
 the $(\Delta I = 1)\,\pi_3-\eta_{8,0}$ mixing and $K^+-K^0$ mass difference, the latter being not amplified by $\omega^{-1}\approx 22$ and thus  negligible relative to the former one in (\ref{6.4}).

 At leading $\ord(p^0,e^2)$,  the $\Omega_{\rm IB}^{(\rm em)}$ component in (\ref{49}) stands for the  electromagnetic isospin-breaking effects ($q_u\not= q_d$)
on the dominant $Q_{4,6}$ {\em on-shell} matrix elements this time via the $(\Delta I=2)$ $\pi^+-\pi^0$ mass difference:
\be\label{51}
\Omega_{\rm IB}^{(\rm em)}(\pi^+-\pi^0)=\left(\frac{\sqrt{2}}{3\omega}\right)\,
\frac{(m_{\pi^0}^2-m_{\pi^+}^2)}{(m_K^2-m_\pi^2)} \approx  -(5.8)\times 10^{-2}.
  \ee
  The subleading $\ord(p^2,e^2)$ corrections including (among others) a genuine  $\Delta I=5/2$ contribution denoted by $f_{5/2}$ in \cite{Cirigliano:2019cpi} tend to
  cancel each other at the percent level, but with huge uncertainties due to renormalization scheme dependence inherent to ChPT. In our approach we safely and consistently   neglect them.

So, combining our main result (\ref{6.4}) for $\Omega_{\rm IB}^{(\rm strong)}$  with (\ref{51}) for $\Omega_{\rm IB}^{(\rm em)}$, we eventually obtain 
\be\label{6.8}
  \boxed{\hat\Omega_{\rm eff}^{(9)}=(29\pm7)\times 10^{-2}}
  \ee
to be compared with the octet result advocated in \cite{Cirigliano:2019cpi}
\be\label{6.9}
  \hat\Omega_{\rm eff}^{(8)}=(17\pm9)\times 10^{-2}.
  \ee
     It is amusing to note that the central value of our 1987 result in (\ref{PhM}), obtained        also in the nonet scheme, practically did not change:
       the increase of $\hat\Omega_{\rm eff}^{(9)}$ through the update of input
       parameters as given in (\ref{6.4}) has been cancelled by the electromagnetic correction in  (\ref{51}) to give the final result in (\ref{6.8}). But the present calculation, performed in the framework of Section~\ref{sec:3}, has a stronger basis than the PhM model and allows the error estimate to be  superior to what was possible at that time.

       In order to appreciate the real impact of the modified value of $\OmHatEff$
       on $\epe$,  we first provide a convenient formula for $\epe$ within the SM that has been recently presented in \cite{Aebischer:2020jto}.
       Consistently with (\ref{eprime}), it reads
\begin{equation}
  \label{eq:semi-num-4}
  \frac{\varepsilon'}{\varepsilon} =
  \im \lambda_t\,\left[
    a_{\rm QCDP}\,(1 - \OmHatEff)  -a_{\rm EWP} \right], \qquad \im \lambda_t=(1.45\pm 0.08)\times 10^{-4}
\end{equation}
with $\OmHatEff$ denoting either $\OmHatEff^{(8)}$ as used
in \cite{Aebischer:2020jto} or $\OmHatEff^{(9)}$ as used by us here below.
As evident from \cite{Aebischer:2020jto}
the first term in (\ref{eq:semi-num-4}) is dominated by the $Q_6$ operator and the second one
by the $Q_8$ operator involving the diagonal quark electric charge matrix $\hat e$ since
\be\label{ANAL}
a_{\rm QCDP}\approx -5.7 + 23.2 \bsi(\mu_0) = 20.0, \qquad  a_{\rm EWP}\approx -2.3 + 9.9 \bei(\mu_2)=4.6\,.
\ee

In order to lower the uncertainties, the numerical coefficients in these expressions have been obtained
by using the hadronic matrix elements of RBC-UKQCD collaboration
at the scales 
\be
\mu_0=4\gev,\qquad \mu_2=3\gev,
\ee
for the isospin amplitudes $A_0$ and $A_2$, respectively. Detailed numerical analysis for other scales is presented in \cite{Aebischer:2020jto}. Here
we confine our discussion to the values of $\bsi(\mu)$ and $\bei(\mu)$, exhibiting their values also for $\mu=1\gev$ to compare 
with the expectations from ChPT and DQCD that work at lower scales than LQCD.

 The values for $\bsi$ and $\bei$,
 extracted from RBC-UKQCD results in  \cite{Abbott:2020hxn} and  \cite{Blum:2015ywa} respectively, are {\cite{Aebischer:2020jto} 
\begin{equation}
  \label{eq:Lbsi}
  \bsi(\mu_0)   = 1.11 \pm 0.20, \qquad 
  \bsi(1\gev)   = 1.49 \pm 0.25,
\end{equation}

\begin{equation}\label{LATB8}
  \bei(\mu_2) 
    = 0.70 \pm 0.04,  \qquad
  \bei(1\gev) 
    = 0.85 \pm 0.05\,.
\end{equation}       

The formula (\ref{eq:semi-num-4}) 
includes NLO QCD corrections to the QCD penguin  (QCDP) contributions and NNLO contributions to electroweak penguins (EWP). We emphasize again that
the IB suppression factor $(1-\hat\Omega_{\rm eff})$  multiplies only the contributions of QCDP operators   while  all EWP contributions, $Q_8$ included, do not involve this {\em scale independent} factor as already assumed in (\ref{DECOMP}).
Such is not necessarily the case in ChPT where, on the basis of
$\text{SU(3)}_L\times\text{SU(3)}_R$ symmetries,
the operator $Q_8$ is ambiguously buried in the effective IB operator
$(U\hat e U^\dagger)_{ds}$ like $Q_4$ and $Q_6$ in $(\partial U\partial  U^\dagger)_{ds}$. It is evident from (\ref{eq:semi-num-4}) that the increased value of  $\hat\Omega_{\rm eff}$
in (\ref{6.8}) implies {a suppression of $\epe$ relative to the value
  presented in \cite{Aebischer:2020jto} on the basis of the octet scheme.

The $\bsi$  value at $\mu=1\gev$ is compatible 
with the estimates from ChPT \cite{Cirigliano:2019ani} while  the one for
$\bei$, with the DQCD estimate in \cite{Buras:2015xba}. As a matter of fact,
the value of $\bei\approx 0.55$ obtained in \cite{Cirigliano:2019ani} by
adding final state interactions (FSI) to the strict large $N$ limit is significantly below the rather
precise LQCD one in (\ref{LATB8}) and casts some doubt on the huge impact of FSI in a partial NLO
estimate within ChPT.

In DQCD FSI have no impact on $\bsi$ in a complete LO estimate \footnote{Contrary to what is claimed in a footnote of \cite{Cirigliano:2019ani}, the absence of FSI in the CP-odd penguin operator at $\ord(p^2,0)+ \ord(p^0,1/N)$ is based on a correct calculation which respects both chiral symmetry and unitarity.} and 
the value of this parameter around $1\gev$ is expected, due to meson evolution, to be below unity
 as required for a smooth matching between hadronic matrix elements and Wilson coefficients. Such a monotonic behaviour with respect to the renormalization scale is indeed observed for the $B_i$-parameters of
 $\Delta S=2$ operators,  as nicely displayed in \cite{Buras:2018lgu}. However,
 the hadronic matrix elements of the left-right
$Q_{5,6}$ penguin operators extracted from the new RBC-UKQCD data taken
at $4~\gev$ correspond to
\be\label{JM1}
\bsi(4~\gev)=1.11\pm0.20,\qquad \frac{\langle Q_5\rangle_0}{\langle Q_6\rangle_0}= 0.31\pm 0.05, \qquad (\text{LQCD}-2020).
\ee

In contrast, at the factorization scale well below $1\gev$ we have
\be\label{JM2}
\bsi(0)=1,\qquad \frac{\langle Q_5\rangle_0}{\langle Q_6\rangle_0}= 0, \qquad
(\text{large-N~limit}).
\ee
Consequently, if confirmed by future LQCD calculations, $\bsi(1\gev)>1$ as
favoured at the $2\sigma$ level by (\ref{eq:Lbsi}) would then imply a rather weird up-down behaviour for  $\bsi(\mu)$
(i.e., $1\to 1.49\to 1.1$) instead of the expected monotonic (decreasing) function with increasing $\mu$ seen for $\bei(\mu)$ with the help of (\ref{LATB8}) and  $\bei(0)=1$.
Such was not the case with the previous LQCD result  \cite{Blum:2015ywa} extracted in \cite{Buras:2015yba}, namely
\be
\bsi(1.5~\gev)=0.57\pm0.19, \qquad (\text{LQCD}-2015).
\ee

On the other hand, the decrease of $\bsi$  from $1~\gev$ to $4~\gev$ amounts in a perturbative regime to a factor of $1.3$. The rate of this decrease in a non-perturbative regime from $\mu\approx 0$ to $\mu=1~\gev$ should be even stronger. Consequently a monotonic decrease of $\bsi$ from very low scales to $4~\gev$ would 
imply, in view of the LQCD result, a value for $\bsi(0)$ by at least a factor of 2 larger than its  large $N$  limit $\bsi(0)=1$.

On the basis of (\ref{JM1}), we also
find intriguing the fact that the naive vacuum-insertion-approximation (VIA)
predictions
\be
\bsi=1, \qquad \frac{\langle Q_5\rangle_0}{\langle Q_6\rangle_0}= \frac{1}{3},
\qquad (\text{VIA})
\ee
are almost fulfilled at a scale as high as $4\gev$.

{ Being back to  $\epe$ in  (\ref{eq:semi-num-4}) after this digression on
the rather controversial low scale dependence of $\bsi$,
the value for  $\hat\Omega^{(9)}_{\rm eff}$ in (\ref{6.8})
implies\footnote{We thank Jason Aebischer and Christoph Bobeth for
  checking this result with more details   given in
  \cite{Aebischer:2020jto}. V2 of the latter paper  uses our result for $ \hat\Omega_{\rm eff}^{(9)}$.}
\be\label{6.10a}
\boxed{(\epe)_\text{SM}^{(9)}=(13.9\pm 5.2)\times 10^{-4}}
\ee
if we trust the new LQCD data displayed in (\ref{ANAL}-\ref{LATB8}).}
It is significantly lower than the value obtained in {\cite{Aebischer:2020jto}}
\be\label{6.10b}
  (\epe)_\text{SM}^{(8)}=(17.4\pm 6.1)\times 10^{-4}\,,
\ee
which used  $\hat\Omega^{(8)}_{\rm eff}$ given in (\ref{6.9}). {Its central value is also   lower by a factor of $1.6$ than the central LQCD value in (\ref{LAT})
  demonstrating that the inclusion of strong isospin-breaking effects in $\epe$
  is very important for the identification of possible NP effects one day.

  As far as the ChPT-2019 and DQCD-2016 predictions for $\epe$ in the SM are concerned, we note
  \begin{itemize}
  \item
    { a total coincidence of (\ref{Pich}) with
    our numerical result} in (\ref{6.10a}) since the former was obtained with
$\hat\Omega^{(8)}_{\rm eff}$ in place of $\hat\Omega^{(9)}_{\rm eff}$ and 
    the values of $\bei\approx 0.55$ and $\im \lambda_t\approx (1.35)\times 10^{-4}$ instead of ours;
  \item
    a still consistent upper bound (\ref{DQCD}) if we only assume a monotonic
    $\bsi(\mu)$, namely $\bsi(1\gev)<1$, instead of $\bsi<\bei$.
  \end{itemize}
     
In the {\em optimal strategy} for the $\epe$ within the SM   proposed recently in  \cite{Aebischer:2019mtr,Buras:2019vik},   $\hat\Omega_{\rm eff}^{(8)}$
  in   (\ref{6.9}) should be replaced
  by $\hat\Omega_{\rm eff}^{(9)}$  in   (\ref{6.8}), the main result of our paper.
  In doing so, the approximate central values in (\ref{Pich}) and (\ref{DQCD})
  read $12\times 10^{-4}$ and $6\times 10^{-4}$, respectively. With the LQCD
  data for $a_{\rm EWP}$, { that disprove ChPT estimate,} the  central value in (\ref{Pich}) would even go
  down to $9\times 10^{-4}$.
\section{Conclusions}\label{sec:7}

An effective $\eta_8-\eta_0$  mixing angle has been exploited to estimate the major impact of the strong isospin-breaking ($m_u\not=m_d$) on $\epe$ via the $\pi_3-\eta_{8,0}$  mixing at work in a full nonet scheme for the pseudoscalars.
As a matter of fact, the lowest-lying $\eta$ pole at $0.5\,\gev$ with its non-negligible  flavour-singlet $\eta_0$ component largely dominates over the $\eta^\prime$  pole sitting
at $1\gev$ as well as over the relevant scalar resonances moved around $1.5\gev$. Such a long-distance effect is yet another challenge for LQCD
final prediction for $\epe$ within the SM.

Taking into account the impact of the electromagnetic isospin-breaking $(q_u\not=q_d)$ dominated by the $\pi^+-\pi^0$ mass difference, we obtain an effective suppression effect of about $40\%$ on the central value of $\epe$ reported
by RBC-UKQCD collaboration in \cite{Abbott:2020hxn}.
The studies of different models until the dusk of the 2010s are listed in Table~3 of \cite{Aebischer:2019mtr}. New activities in this direction, including correlations with other observables, are expected at the dawn of the 2020s.

\section*{Acknowledgements}
We thank Jason Aebischer and Christoph Bobeth for discussions.
This research was supported by the Excellence Cluster ORIGINS,
funded by the Deutsche Forschungsgemeinschaft (DFG, German Research Foundation) under Germany´s Excellence Strategy – EXC-2094 – 390783311.

\renewcommand{\refname}{R\lowercase{eferences}}

\addcontentsline{toc}{section}{References}

\bibliographystyle{JHEP}
\bibliography{Bookallrefs}
\end{document}